\documentclass[aps,showpacs,amsmath,amssymb]{revtex4}
\usepackage{amsmath,amssymb,amsthm}
\usepackage[colorlinks,urlcolor=blue,anchorcolor=blue,citecolor=blue]{hyperref}

\begin{document}

\title{Remote implementation of partially unknown operations and its entanglement costs}
\author{Shu-Hui Luo\footnote{lsh1990@mail.ustc.edu.cn},An-Min Wang\footnote{amwang@ustc.edu.cn}}
\affiliation{Quantum Theory Group, Department of Modern Physics\\
University of Science and Technology of China, Hefei 230026, People
Republic of China}

\begin{abstract}
We present the generalized version of Wang's protocol[A.M.Wang, Phys.Rev.A 74,032317 (2006)] for the remote implementation(sometimes referred to as quantum remote control) of partially unknown quantum operations. The protocol only requires no more than half of the entanglements used in Bidirectional Quantum State Teleportation. We also propose a protocol for another form of quantum remote control. It can remotely implement a unitary operation which is a combination of the projective representations of a group. Moreover, we prove that the Schmidt rank of the entanglements cannot not be less than the number of controlled parameters of the operations, which for the first time gives a lower bound on entanglement costs in remote implementation of quantum operations.
\end{abstract}
\maketitle

\section{INTRODUCTION}
In the construction of a quantum computer, it is difficult to maintain all qubits in a single processor due to decoherence. One alternative way is to build it as a multiprocessor device, that is to say, each processor contains only a few qubits. Evidently, such a "distributed quantum computer"\cite{Cirac1999} requires the remote implementation of quantum operations(RIO, sometimes also referred to as quantum remote control) since each processor can only perform limited local operations or it may not know all the information of the operations. Besides, RIO may also play important roles in distributed quantum computation, large scale quantum simulation, quantum programs or other remote quantum information processing tasks.

Without physically moving the qubits around, we can remotely implement operations using only local quantum operations and classical communications(LOCC) and prior entanglements. One straightforward way is resorting to Bidirectional Quantum State Teleportation(BQST)\cite{Bennett1993}, where we teleport all the qubits involved to one party, and teleport them back after the desired operation is performed. And it is proved that when the operation is completely unknown, we can only rely on BQST \cite{Huelga2001}. BQST requires two rounds of teleportation. Indeed, it sets the upper bound on the resources needed for RIO.

As entanglements are valuable resources in quantum information and quantum computation, which are difficult to create and maintain, we should seek methods to save entanglements. Fortunately, when the operation falls into some restricted sets, we are able to remotely implement it using fewer entanglements via some protocols\cite{Huelga2002,Wang2006,Wang2007,Zhao2007} than via BQST. Some experiments have been demonstrated\cite{Xiang2005,Huelga2005}.

Reference \cite{Huelga2002} proposed the HPV protocol for quantum remote control of diagonal or anti-diagonal one-qubit operations. Reference \cite{Wang2006} extended the HPV protocol to the Wang's protocol for the remote implementation of partially unknown multiqubits operations where there is only one nonzero element in every row or every column of the operations. By saying "partially unknown", we mean that Alice, the party that holds the quantum state to be operated on, does not know all the information of the remote operations. In the Wang's protocol, Alice only knows the structure of the operations, but not the nonzero elements of the operations. Reference \cite{Wang2007} presented the protocols for combined and controlled remote implementations of partially unknown quantum operations of multiqubits using Greenberger-Horne-Zeilinger states. Reference \cite{Zhao2007} presented a hybrid protocol of remote implementations of quantum operations.

However, all the previous references did not give the minimum entanglement costs in RIO, even a lower bound. So we are curious about what is the necessary entanglement costs in RIO and whether there is a lower bound for entanglement costs in RIO. Recently we obtained a conclusion that the Schmidt rank(defined in \cite{QCQI}) of the entanglements cannot not be less than the number of controlled parameters of the operations, which for the first time provides a general lower bound on the required entanglement resource and gives a criterion to assess protocols for RIO. However, we haven't proved that the entanglement state should be maximally entangled, which we think should be the case.

Local implementation of nonlocal unitaries\cite{Eisert2000,Huang2004,Zhao2008,Yu2010} is a different issue. In fact, there are profound connections between local implementation of nonlocal unitaries and RIO. Reference \cite{Yu2010} proposed a protocol for implementing nonlocal controlled unitaries of the form $\mathcal{U}=\sum _{j=0}^{N-1} P_j\otimes V_j$ where the $P_j$'s form a projective decomposition of the identity on $\mathcal{H}_A$, while the $V_j$'s are arbitrary unitaries on $\mathcal{H}_B$. Reference \cite{Yu2010} also presented a protocol for local implementation of nonlocal unitaries of the group decomposition form $\mathcal{U}=\sum _{f\in G} U(f)\otimes W(f)$ where the unitary operators $U(f)$'s form a finite-dimensional projective representation of a group $G$.

Inspired by Reference \cite{Wang2006,Yu2010}, we figure out the generalized version of Wang's protocol\cite{Wang2006} and a protocol for remote implementation of quantum operations which are combinations of the projective representations of groups. The former is able to remotely implement operations of the form
\begin{equation}
\mathcal{U}=\sum _{i=0}^{N-1} c_iA_i,
\end{equation}
where $c_i$'s are $N$ arbitrary complex coefficients with modulus unity and $A_i$'s satisfy $A_i^{\dagger}A_j=0,\text{ for }i\neq j$. The latter is able to remotely implement operations of the form
\begin{equation}
\mathcal{U}=\sum _{f\in G} c(f)U(f)
\end{equation}
where the unitary operators $U(f)$'s form a finite-dimensional projective representation of a group $G$ and $c(f)$'s are complex coefficients determined by $\mathcal{U}$.

Both of the protocols can remotely implement operations of certain forms using less entanglements than BQST. And they are indeed different kinds of quantum remote control. Previous protocols are quantum remote control similar to the generalized version of Wang's protocol. So it is worthwhile to present the Wang's protocol in a general way. But the protocol for RIO of group form is a different kind of quantum remote control. The form of remote operations it implements and the local operations it needs are different from those in the generalized version of Wang's protocol. Thus the protocol for RIO of group form is a good supplement to quantum remote control. There may still be other forms of quantum remote control. In a word, both the protocols we present in this paper would enhance the power of RIO and extend the applications of RIO.

The remainder of this paper is organized as follows. In Sec.\ref{sec2}, we present the generalized version of Wang's protocol for RIO. Section \ref{sec3} presents another protocol for RIO of the group form. Section \ref{sec4} proves that the Schmidt rank of entanglements cannot not be less than the number of controlled parameters of the operations. In Sec.\ref{sec5}, we conclude our results.

\section{A PROTOCOL OF RIO} \label{sec2}
Reference \cite{Huelga2002} presented a protocol for the remote implementation of quantum operations on a single qubit, where the operations are diagonal or anti-diagonal. Reference \cite{Wang2006} presented the Wang's protocol for the remote implementation of partially unknown operations
\begin{equation}
U(x)=\sum _{i=0}^{2^N-1} e^{\text{i$\phi $}_i}\left|p_i(x),D\right\rangle \langle i,D|
\end{equation}
on $N$ qubits, where $\phi_i$'s are $2^N$ real phases and $D$ indicates the decimal system, i.e., $|0,D\rangle =|00\text{$\ldots $0}\rangle ,|1,D\rangle =|00\text{$\ldots $1}\rangle ,$ and $\left|2^N-1,D\right\rangle =|11\text{$\ldots $1}\rangle$, etc. $p(x)=\left\{p_0(x),p_1(x),\ldots ,p_{2^N-1}(x)\right\}$ is a permutation of the list $\left\{0,1,\ldots ,2^N-1\right\}$, where $x=1,2,\ldots ,2^N!$ maps all the permutations. This protocol uses $N$ Bell states.

Now we present the generalized version of Wang's protocol for the remote implementation of operations in the form
\begin{equation}
\mathcal{U}=\sum _{i=0}^{N-1} c_iA_i
\end{equation}
where $c_i$'s are $N$ arbitrary complex coefficients with modulus unity and $A_i$'s satisfy $A_i^{\dagger}A_j=0,\text{ for }i\neq j$. Note that unlike the Wang's protocol, $N$ here, the number of controlled parameters, has no relationship with the dimensionality of the Hilbert Space. Due to unitarity of $\mathcal{U}$, it can be proved that
\begin{equation}
A_i=\sum _{j=1}^{r_i} c_i|v_j^{(i)}\rangle \langle u_j^{(i)}|
\end{equation}
where $\left\{\left|v_j\right\rangle \right\}$ and $\left\{\left|u_j\right\rangle \right\}$ are two full sets of mutually orthonormal vectors of the Hilbert Space and the superscript $(i)$ indicates that a certain vector belongs to and only belongs to a certain $A_i$, so $A_i$'s satisfy $A_i^{\dagger}A_j=0,\text{ for }i\neq j$. And $r_i$ is the rank of $A_i$. The operations in Wang's protocol are special cases of this form when the ranks of $A_i$'s equal to one.

We want to implement the operation on $\mathcal{H}_A$. Let the initial state of $\mathcal{H}_A$ be $|\Psi\rangle_A$. The entanglement resource is
\begin{equation}
|\Phi \rangle _{\text{ab}}=\frac{1}{\sqrt{N}}\sum _{k=0}^{N-1} |k\rangle \otimes |k\rangle,
\end{equation}
where the dimension of $\mathcal{H}_a$ or $\mathcal{H}_b$ is $N$. For a given $N$ define the $X$ gate such that
\begin{equation}
X|k\rangle =|k-1\rangle.
\end{equation}
Here, subtraction should be understood as mod $N$. The protocol has the following steps.

\paragraph*{step 1}
Alice performs
\begin{equation}
\mathcal{P}=\sum _{i=0}^{N-1} P_i\otimes X^i, \text{ where }P_i=\sum _{j=1}^{r_i} |u_j^{(i)}\rangle \langle u_j^{(i)}|
\end{equation}
acts on $\mathcal{H}_A$, and $X^i$ means $X$ to the power $i$, which acts on $\mathcal{H}_a$. After this step the state of the combined system becomes
\begin{equation}
\sum _{i=0}^{N-1} P_i|\Psi \rangle _A\otimes \frac{1}{\sqrt{N}}\sum _{k=0}^{N-1} |k-i\rangle _a\otimes |k\rangle _b.
\end{equation}

\paragraph*{step 2}
Alice performs a measurement on $\mathcal{H}_a$ in the computational basis. The measurement result $l$ is sent to Bob. Bob then performs $X^l$ on $\mathcal{H}_b$. The state of the system is now
\begin{equation}
\sum _{i=0}^{N-1} P_i|\Psi \rangle _A\otimes |i\rangle _b.
\end{equation}

\paragraph*{step 3}
Bob performs $C=\sum _{i=0}^{N-1} c_i|i\rangle _b\langle i|$
on $\mathcal{H}_b$. The state of the system is
\begin{equation}
\sum _{i=0}^{N-1} c_iP_i|\Psi \rangle _A\otimes |i\rangle _b.
\end{equation}

\paragraph*{step 4}
Bob performs a Fourier transform
\begin{equation}
F=\frac{1}{\sqrt{N}}\sum _{m,j=0}^{N-1} e^{2\text{$\pi $imj}/N}|m\rangle \langle j|
\end{equation}
on $\mathcal{H}_b$ and then measures $\mathcal{H}_b$ in the computational basis. The measurement result $m$ is sent to Alice. The state of the system becomes
\begin{equation}
\sum _{j=0}^{N-1} e^{2\text{$\pi $imj}/N}c_jP_j|\Psi \rangle _A\otimes |m\rangle _b.
\end{equation}

\paragraph*{step 5}
Alice performs
\begin{equation}
\mathcal{R}_m=\sum _{j=0}^{N-1} \sum _{k=1}^{r_j} e^{-2 \text{$\pi $imj}/N}|v_k^{(j)}\rangle \langle u_k^{(j)}|
\end{equation}
on $\mathcal{H}_A$. It completes the implementation of the operation. The final state is exactly
\begin{equation}
\mathcal{U}|\Psi \rangle _A=\sum _{j=0}^{N-1} c_jA_j|\Psi \rangle _A=\sum _{j=0}^{N-1} \sum _{k=1}^{r_j} c_j|v_k^{(j)}\rangle \langle u_k^{(j)}|\Psi \rangle _A.
\end{equation}

It can be proved that operations of this form are all we can implement using such a protocol. Notice that $c_i$'s can be chosen arbitrarily when $A_i$'s are given. Besides, $c_i$'s may be unknown to Alice, which keeps her from implementing the operation locally and obliges her to resort to RIO. Here lies the essence of quantum remote control\cite{Huelga2001,Huelga2002}. By such a protocol, Bob can apply a controlled and private operation on Alice's quantum state.

In fact, if Alice is able to locally implement any operation, by this protocol any unitary can be remotely implemented while Alice does not know all the information of the unitary. Because any unitary can be decomposed to a diagonal matrix with a unitary on each side by Singular Value Decomposition(SVD). For example $\mathcal{U}=\sum _{i=0}^{N-1} |u_i\rangle d_i \langle v_i|=\sum _{i=0}^{N-1} |u_i\rangle \langle i| \sum _{j=0}^{N-1} |j\rangle d_j \langle j| \sum _{k=0}^{N-1} |k\rangle \langle v_k|=udv$. If Bob wants to remotely implement a certain unitary $\mathcal{U}$ on Alice's quantum state while keeping Alice from knowing all the information of $\mathcal{U}$, he can first calculate the SVD of $\mathcal{U}=udv$ and tell Alice $u$ and $v$. Then Alice performs $v$ on her state. After that Bob remotely implements $d$ using the above protocol. Finally Alice performs $u$. By this mean, $\mathcal{U}$ is remotely implemented while Alice does not know the elements of $d$. This method is nontrivial since if we use BQST instead, the entanglement costs would double.

But notice, in the above process Alice should have the devices to perform $u$ and $v$. So if Alice and Bob want to remotely implement any unitary, Alice should have the devices to perform any local operation. Hence given limited devices, Alice and Bob can only remotely implement operations in some restricted sets. To enhance the power of RIO, we will present another protocol in the next section which can achieve a different form of RIO.

\section{RIO OF GROUP FORM} \label{sec3}
Reference \cite{Yu2010} presented a protocol for local implementation of nonlocal unitaries of the group decomposition form
\begin{equation}
\mathcal{U}=\sum _{f\in G} U(f)\otimes W(f)
\end{equation}
where the unitary operators $U(f)$'s form a finite-dimensional projective representation of a group $G$. By saying projective representation, that means
\begin{equation}
U(f) U(g)= \mu (f,g)U(f g)
\end{equation}
where $\mu(f,g)$'s are complex numbers constituting a factor system. Because of unitarity condition, $\mu(f,g)$'s are of modulus one. Using
\begin{equation}
U(g)=U(f^{-1})U(f)U(g)= \mu (f,g)U(f^{-1})U(f g)=\mu (f,g)\mu (f^{-1},f g)U(g),
\end{equation}
we have $\mu (f,g)\mu (f^{-1},f g)=1$. Hence, $\mu \left(h^{-1},f\right)\mu \left(h,h^{-1}f\right)=1$. We will use this identity later.

In the following passages we will demonstrate the protocol to remotely implement operations of the form
\begin{equation}
\mathcal{U}=\sum _{f\in G} c(f)U(f)
\end{equation}
where the unitary operators $U(f)$'s form a finite-dimensional projective representation of a group $G$ and $c(f)$'s are controlled complex coefficients.

Before presenting the protocol, we first make some reasoning. The reasoning was motivated by the discussion in Part.II.B of Ref.\cite{Yu2010}. Suppose we want to remotely implement an operation $\mathcal{U}=\sum _{i=0}^{N-1} c_iU_i$ on $\mathcal{H}_A$. And assume that the first two steps in Sec. \ref{sec2} are necessary with $P_i$'s being undefined. After the first two steps, we arrive at
\begin{equation}
\sum _{i=0}^{N-1} P_i|\Psi \rangle _A\otimes |i\rangle _b.
\end{equation}
Then Bob performs an operation $M$ on $\mathcal{H}_b$ and then measures $\mathcal{H}_b$ in the computational basis. The measurement result $m$ is sent to Alice. Alice then performs a corresponding recovery operation $\mathcal{R}_m$ on $\mathcal{H}_A$. The final state of the system becomes
\begin{equation}
\sum _{i=0}^{N-1} \langle m|M|i\rangle R_mP_i|\Psi \rangle _A\otimes |m\rangle _b.
\end{equation}
So if
\begin{equation}
\sum _{i=0}^{N-1} \langle m|M|i\rangle R_mP_i=\sum _{i=0}^{N-1} c_iU_i,
\end{equation}
we have successfully applied the operation.

Particularly, define
\begin{equation}
U_i=U\left(g_i\right), R_m=U\left(g_m^{-1}\right), P_i=U\left(g_i\right), \langle m|M|i\rangle =\mu \left(g_m^{-1},g_i\right)^{-1}c\left(g_m^{-1}g_i\right)
\end{equation}
where $g_i$'s are elements of a group $G$ labeled by $i$ and $U(g_i)$'s are their projective representations. Thanks to the Rearrangement Theorem in group theory,
\begin{equation}
\sum _{i=0}^{N-1} c\left(g_m^{-1}g_i\right)U\left(g_m^{-1}g_i\right)=\sum _{i=0}^{N-1} c\left(g_i\right)U\left(g_i\right).
\end{equation}
Thus, we can successfully implement $\mathcal{U}$ by defining the operations as above.

The protocol follows from five steps.
\paragraph*{step 1}
Alice performs $\mathcal{P}=\sum _{f\in G} U(f)\otimes |f\rangle _a\langle f|$ where $|f\rangle_a$ are orthonormal basis on $\mathcal{H}_a$. After this step the state of the combined system becomes
\begin{equation}
\frac{1}{\sqrt{|G|}}\sum _{f\in G} U(f)|\Psi \rangle _A\otimes |f\rangle _a|f\rangle _b.
\end{equation}

\paragraph*{step 2}
Alice performs $F$ on $\mathcal{H}_a$ and then makes a measurement. The measurement result $g$ is sent to Bob. The state of the system is
\begin{equation}
\sum _{f\in G} U(f)|\Psi \rangle _A\otimes \langle g|F|f\rangle |g\rangle _a|f\rangle _b.
\end{equation}

\paragraph*{step 3}
Bob then performs $Z(g)$ on $\mathcal{H}_b$. $Z(g)$ is defined as
\begin{equation}
Z(g)|f\rangle =\frac{1}{\sqrt{|G|}}\langle g|F|f\rangle ^{-1}|f\rangle.
\end{equation}
The state of the system is now
\begin{equation}
\frac{1}{\sqrt{|G|}}\sum _{f\in G} U(f)|\Psi \rangle _A\otimes |g\rangle _a|f\rangle _b.
\end{equation}

\paragraph*{step 4}
Bob performs $M$ on $\mathcal{H}_b$. $M$ is defined as
\begin{equation}
M=\sum _{f\in G} c(f)R(f), R(f)=\sum _{g\in G} \mu (g,f)|g\rangle \langle gf|.
\end{equation}
Then Bob performs a measurement on $\mathcal{H}_b$. The measurement result $h$ is sent to Alice. The state of the system becomes
\begin{equation}
\sum _{f\in G} U(f)|\Psi \rangle _A\otimes c\left(h^{-1}f\right)\mu \left(h,h^{-1}f\right)|g\rangle _a|h\rangle _b.
\end{equation}

\paragraph*{step 5}
Alice performs $U(h^{-1})$ on $\mathcal{H}_A$. This completes the protocol. With $\mu \left(h^{-1},f\right)\mu \left(h,h^{-1}f\right)=1$, we arrive at
\begin{equation}
\mathcal{U}|\Psi \rangle _A\otimes |g\rangle _a|h\rangle _b=\sum _{f\in G} c(f)U(f)|\Psi \rangle _A\otimes |g\rangle _a|h\rangle _b.
\end{equation}

It is easy to prove that this protocol is as general as BQST. The proof is similar to that in Part.V.A of Ref.\cite{Yu2010}. And for a given projective representation of $G$ and a given $\mathcal{U}$, $c(f)$'s are determined as
\begin{equation}\label{1}
c(f)=\sum _{\lambda =1}^{\kappa } \frac{d_{\lambda }}{N}\sum _{j,k=1}^{d_{\lambda }} \left[D_{\text{jk}}^{(\lambda )}(f)\right]^*\mathcal{R}_{\text{jk}}^{(\lambda )}
\end{equation}
where the notation is similar to that in Part.IV.D of Ref.\cite{Yu2010}. For a given group, there are $\kappa$ inequivalent unitary irreducible representations $\{D^{(\lambda)}(f)\}$ labeled by $\lambda$, where $D^{(\lambda)}(f)$ is a $d_\lambda \times d_\lambda$ matrix. And $\sum _{\lambda=1}^{\kappa} d_{\lambda}^2=|G|=N$. In a certain basis of $\mathcal{H}_A$, $U(f)$ can be expressed in a block diagonal form
\begin{equation}
U(f)=\bigoplus _{\lambda=1}^{\kappa} D^{(\lambda)}(f).
\end{equation}
Thus in that basis, $\mathcal{U}$ can also be expressed in a block diagonal form
\begin{equation}
\mathcal{U}=\sum _{f\in G} c(f)U(f)=\bigoplus _{\lambda=1}^{\kappa} \mathcal{R}^{(\lambda)}.
\end{equation}
For simplicity, we are only talking in the situation in which the representation $U(f)$ contains each inequivalent irreducible representation exactly once. Please refer to Part.IV.C of Ref.\cite{Yu2010} for further discussions. As the Double Unitary protocol in Part.IV.D of Ref.\cite{Yu2010} is valid, our protocol is valid either.

Last but not least, Alice only needs the devices to perform $\mathcal{P}$, $F$ and $U(f)$'s in the protocol. Given these devices, any combination of $U(f)$'s can be remotely implemented as long as the operation is unitary.

\section{ENTANGLEMENT COSTS} \label{sec4}
Alice has no information of $c_i$'s. Hence, the information should be transmitted from Bob to Alice. By what means? Entanglements.
$A_i$'s and $c_i$'s are coupled with the aid of entanglements.

For heuristic reason, first go through the protocol in Sec.\ref{sec2}. We will use a diagrammatic method\cite{Cohen2007} to express the process.

Let the initial state of the system be $|\Psi \rangle _A\otimes \overset{2}{\sum _{i=0} }\frac{1}{\sqrt{3}}|i\rangle _a|i\rangle _b$. With the notation in Part.II.B of Ref.\cite{Yu2010}, it can be expressed as such a matrix:
\begin{table}[!ht]
\centering
\begin{tabular}{c|c|c|c|}
  a$\backslash$b & 0 & 1 & 2 \\  \hline
  0 & $\frac{1}{\sqrt{3}}|\Psi \rangle_A$ & 0 & 0 \\ \hline
  1 & 0 & $\frac{1}{\sqrt{3}}|\Psi \rangle_A$ & 0 \\ \hline
  2 & 0 & 0 & $\frac{1}{\sqrt{3}}|\Psi \rangle_A$ \\ \hline
\end{tabular}
\end{table}\\
Here, the state of $\mathcal{H}_a$ is expressed as a column and the state of $\mathcal{H}_b$ is expressed as a row.

We want to implement
\begin{equation}
\mathcal{U}|\Psi \rangle=\sum _{i=0}^2 \sum _{j=1}^{r_i}c_i|v_j^{(i)}\rangle \langle u_j^{(i)}|\Psi \rangle=\sum _{i=0}^2 c_iA_i|\Psi \rangle_A=
\left(
\begin{array}{ccc}
 A_0 & A_1 & A_2
\end{array}
\right)\left(
\begin{array}{ccc}
 |\Psi \rangle_A  &  &  \\
  & |\Psi \rangle_A  &  \\
  &  & |\Psi \rangle_A
\end{array}
\right)\left(
\begin{array}{c}
 c_0 \\
 c_1 \\
 c_2
\end{array}
\right).
\end{equation}

\paragraph*{step 1}
Alice performs
\begin{equation}
\mathcal{P}=\sum _{i=0}^2 P_i\otimes X^i=\left(
\begin{array}{ccc}
 P_0 & P_1 & P_2 \\
 P_2 & P_0 & P_1 \\
 P_1 & P_2 & P_0
\end{array}
\right)\text{, where }P_i=\sum _{j=1}^{r_i}|u_j^{(i)}\rangle \langle u_j^{(i)}|.
\end{equation}
After this step the state of the combined system becomes
\begin{equation}
\sum _{i=0}^{2} P_i|\Psi \rangle _A\otimes \frac{1}{\sqrt{3}}\sum _{k=0}^{2} |k-i\rangle _a\otimes |k\rangle _b=\frac{1}{\sqrt{3}}
\left(
\begin{array}{ccc}
 P_0|\Psi \rangle_A & P_1|\Psi \rangle_A & P_2|\Psi \rangle_A \\
 P_2|\Psi \rangle_A & P_0|\Psi \rangle_A & P_1|\Psi \rangle_A \\
 P_1|\Psi \rangle_A & P_2|\Psi \rangle_A & P_0|\Psi \rangle_A
\end{array}
\right).
\end{equation}

\paragraph*{step 2}
Alice performs a measurement on $\mathcal{H}_a$ in the computational basis. The measurement result $l=1$ is sent to Bob. Bob then performs $X$ on $\mathcal{H}_b$. The state of the system is now
\begin{equation}
\left(
\begin{array}{ccc}
 P_2|\Psi \rangle_A & P_0|\Psi \rangle_A & P_1|\Psi \rangle_A \\
\end{array}
\right)
\left(
\begin{array}{ccc}
 0 & 0 & 1 \\
 1 & 0 & 0 \\
 0 & 1 & 0
\end{array}
\right)=
\left(
\begin{array}{ccc}
 P_0|\Psi \rangle_A & P_1|\Psi \rangle_A & P_2|\Psi \rangle_A
\end{array}
\right).
\end{equation}
Because the state of $\mathcal{H}_b$ is expressed as a row, we use the transpose form of $X$. And we multiply $\sqrt{3}$ to preserve unitarity after performing a measurement.

\paragraph*{step 3}
Bob performs
\begin{equation}
C=\sum _{i=0}^{2} c_i|i\rangle _b\langle i|=
\left(
\begin{array}{ccc}
 c_0 &     &  \\
     & c_1 &  \\
     &     & c_2
\end{array}
\right)
\end{equation}
on $\mathcal{H}_b$. The state of the system is
\begin{equation}
\sum _{i=0}^{2} c_iP_i|\Psi \rangle _A\otimes |i\rangle _b=
\left(\begin{array}{ccc}
 c_0P_0|\Psi \rangle_A & c_1P_1|\Psi \rangle_A & c_2P_2|\Psi \rangle_A
\end{array}
\right).
\end{equation}

\paragraph*{step 4}
Bob performs a Fourier transform
\begin{equation}
F=\frac{1}{\sqrt{3}}\sum _{m,j=0}^{2} e^{2\text{$\pi $imj}/3}|m\rangle \langle j|=
\frac{1}{\sqrt{3}}\left(
\begin{array}{ccc}
 1 & 1 & 1 \\
 1 & e^{\frac{2\text{$\pi $i}}{3}} & e^{\frac{4\text{$\pi $i}}{3}} \\
 1 & e^{\frac{4\text{$\pi $i}}{3}} & e^{\frac{8\text{$\pi $i}}{3}}
\end{array}
\right)
\end{equation}
on $\mathcal{H}_b$.
\begin{equation}
\sum _{i=0}^{2} c_iP_i|\Psi \rangle _A\otimes F|i\rangle _b=
\left(\begin{array}{ccc}
 c_0P_0|\Psi \rangle_A & c_1P_1|\Psi \rangle_A & c_2P_2|\Psi \rangle_A
\end{array}
\right)
\frac{1}{\sqrt{3}}\left(
\begin{array}{ccc}
 1 & 1 & 1 \\
 1 & e^{\frac{2\text{$\pi $i}}{3}} & e^{\frac{4\text{$\pi $i}}{3}} \\
 1 & e^{\frac{4\text{$\pi $i}}{3}} & e^{\frac{8\text{$\pi $i}}{3}}
\end{array}
\right)
\end{equation}
Then measures $\mathcal{H}_b$ in the computational basis. The measurement result $m=2$ is sent to Alice. The state of the system becomes
\begin{equation}
\sum _{j=0}^{2} e^{4\text{$\pi $ij}/3}c_jP_j|\Psi \rangle _A=c_0P_0|\Psi \rangle_A+e^{\frac{4\text{$\pi $i}}{3}}c_1P_1|\Psi \rangle_A+e^{\frac{8\text{$\pi $i}}{3}}c_2P_2|\Psi \rangle_A.
\end{equation}

\paragraph*{step 5}
Alice performs
\begin{equation}
\mathcal{R}_m=\sum _{j=0}^{2} \sum _{k=1}^{r_j} e^{-2 \text{$\pi $imj}/3}|v_k^{(j)}\rangle \langle u_k^{(j)}|
\end{equation}
on $\mathcal{H}_A$. It completes the remote implementation of the operation. The final state is exactly
\begin{equation}
\mathcal{U}|\Psi \rangle _A=\sum _{i=0}^{2} c_iA_i|\Psi \rangle _A=\sum _{i=0}^{2} \sum _{j=1}^{r_i} c_i|v_j^{(i)}\rangle \langle u_j^{(i)}|\Psi \rangle _A.
\end{equation}

Indeed, the implementation can be summarized as follows. Assume that the measurement result of $\mathcal{H}_a$ is $l$ and that of $\mathcal{H}_b$ is $m$. Obviously, Bob's action is subject to $l$. And the consequence of the measurement on $\mathcal{H}_a$ is to pick a row out of Alice's operation matrix, let it be $[\mathcal{P}_{ij}]$, while the consequence of the measurement on $\mathcal{H}_b$ is to pick a column out of Bob's operation matrix, let it be $[M^{(l)}_{ij}]^T$. In the end, Alice performs a recovery operation $\mathcal{R}_m$ on $\mathcal{H}_A$. The whole processing can be expressed as
\begin{equation*}
\mathcal{U}|\Psi \rangle _A=\sqrt{n}\mathcal{R}_m
\left(
\begin{array}{cccc}
 \mathcal{P}_{l 0} & \mathcal{P}_{l 1} & \cdots  & \mathcal{P}_{l (n-1)}
\end{array}
\right)\left(
\begin{array}{cccc}
 |\Psi \rangle _A & 0 & \cdots  & 0 \\
 0 & |\Psi \rangle _A & \cdots  & 0 \\
 \vdots  & \vdots  & \ddots & \vdots  \\
 0 & 0 & \cdots  & |\Psi \rangle _A
\end{array}
\right)\left(
\begin{array}{c}
 M^{(l)}{}_{m 0} \\
 M^{(l)}{}_{m 1} \\
 \vdots  \\
 M^{(l)}{}_{m (n-1)}
\end{array}
\right)
\end{equation*}

\begin{equation}=
\left(
\begin{array}{cccc}
 A_0 & A_1 & \cdots  & A_{n-1}
\end{array}
\right)
\left(
\begin{array}{cccc}
 |\Psi \rangle _A & 0 & \cdots  & 0 \\
 0 & |\Psi \rangle _A & \cdots  & 0 \\
 \vdots  & \vdots  & \ddots & \vdots  \\
 0 & 0 & \cdots  & |\Psi \rangle _A
\end{array}
\right)
\left(
\begin{array}{c}
 c_0 \\
 c_1 \\
 \vdots  \\
 c_{n-1}
\end{array}
\right).
\end{equation}
Note, here the dimensionality of $\mathcal{H}_a$ or $\mathcal{H}_b$, or the Schmidt rank of entanglements, is equal to $n$, the number of $c_i$'s, the controlled parameters. And the entanglement state is maximally entangled. Is it possible that $d$, the dimensionality of $\mathcal{H}_a$ or $\mathcal{H}_b$, is smaller than $n$? Or is it possible that the entanglement state is partially entangled?

Suppose it is possible. The entanglement state is $\sum _{i=0}^{d-1} \frac{1}{\sqrt{d}}h_i|i\rangle _a|i\rangle _b$(With Schmidt decomposition, we can always expressed entangled state in a diagonal form in a suitable basis). We have
\begin{equation}
\mathcal{U}|\Psi \rangle _A=\sum _{i=0}^{n-1} c_iA_i=\sqrt{d}\mathcal{R}_m\sum _{j=0}^{d-1} h_jM^{(l)}{}_{m j}\mathcal{P}_{l j}.
\end{equation}
Since Alice has no knowledge of $c_i$'s, when $c_i$'s change and $A_i$'s remain the same, Alice's action on $\mathcal{H}_A$, $\mathcal{P}$, should stay the same. Because $c_i$'s are $n$ arbitrary phase factors of modulus unity, their degree of freedom is $n$. When $c_i$'s change and $A_i$'s and $\mathcal{P}$ remain the same, the equation always holds. Thus, $M^{(l)}{}_{m j}$'s should be some linear combinations of $c_i$'s. Just define
\begin{equation}
M^{(l)}{}_{m j}=\overset{n-1}{\sum _{i=0} }q^{(l m)}{}_{j i}c_i.
\end{equation}
Then we find
\begin{equation}\label{2}
A_i=\sqrt{d}\mathcal{R}_m\overset{d-1}{\sum _{j=0} }h_jq^{(l m)}{}_{j i}\mathcal{P}_{l j}.
\end{equation}
$\mathcal{P}_{l j}(j=0,1,\cdots ,d-1)$ are at most $d$ linearly independent operators. From (\ref{2}), we can see $A_i$'s are at most $d$ linearly independent operators either because the rank of $[h_jq^{(l m)}{}_{j i}]$ may not exceed $d$. However, $A_i$'s are $n$ linearly independent operators by definition. So we come to the conclusion that the dimensionality of the entanglement resource, or the Schmidt rank of entanglements, cannot be smaller than the number of $c_i$'s, the controlled parameters.

However, we are still unable to answer whether the entanglement resource should be maximally entangled. We believe the answer is yes.

Though the coefficients $c(f)$'s in the protocol of Sec.\ref{sec3} are subject to (\ref{1}), it can be easily proved that their degree of freedom is also equal to $|G|$, the number of elements in the group, since $n^2$ real parameters are needed to determine an $n\times n$ unitary matrix. We can prove that the Schmidt rank of entanglements required by that protocol cannot be less than $|G|$ similarly.

\section{CONCLUSION} \label{sec5}
We present the generalized Wang's protocol for the remote implementation(remote control) of partially unknown quantum operations. We also propose the quantum remote control of group form. The protocols enhance the power of RIO and extend the applications of RIO. Then we prove that the Schmidt rank of the entanglement state cannot be less than the number of controlled parameters, which provides a lower bound for entanglement costs in RIO. But we are still unable to prove that the entanglement resource should be maximally entangled while previous protocols all require a maximally entangled state. This will be left for future study.

Our work analyzes the protocols for remote implementation of partially unknown quantum operations in detail and will provide some clues for new protocols, such as protocols for other forms of quantum remote control. Our work gives the necessary Schmidt rank of the entanglement resource for the first time. It provides clues for the minimum entanglement costs for RIO and gives a standard to evaluate previous protocols or future protocols. Since RIO has important applications in Quantum Information and Quantum Computation, our results are nontrivial. Future study should be cast to proposing protocols for other forms of quantum remote control and giving its minimum entanglement costs.

\section*{ACKNOWLEDGEMENTS}
This work was supported by the National Natural Science Foundation of China under Grant No 10975125.

\bibliography{Reference}

\end{document}